\documentclass[12pt,preprint]{aastex}

\newcommand{\be}{\begin{equation}}
\newcommand{\ee}{\end{equation}}

\shorttitle{Compact radio structure of radio-loud NLS1s}
\shortauthors{M. F. Gu and Y. J. Chen}

\begin{document}


\title{The compact radio structure of radio-loud narrow line Seyfert 1 galaxies}


\author{Minfeng Gu\altaffilmark{1} and Yongjun Chen}
\affil{Key Laboratory for Research in Galaxies and Cosmology,
Shanghai Astronomical Observatory, Chinese Academy of Sciences, 80
Nandan Road, Shanghai 200030, China}

\email{gumf@shao.ac.cn}


\altaffiltext{1}{Department of Physics, University of California,
Santa Barbara, CA 93106, USA}


\begin{abstract}

We present the compact radio structure of three radio-loud narrow
line Seyfert 1 galaxies from VLBA archive data at 2.3, 5 and 8.4
GHz. In RXS J16290+4007, the radio structure is mostly unresolved.
The combination of compact radio structure, high brightness
temperature and inverted spectrum between simultaneous 2.3 and 8.4
GHz, strongly favors jet relativistic beaming. Combining with the
VLBI data at 1.6 and 8.4 GHz from literatures, we argued that RXS
J16333+4718 may also harbor a relativistic jet, with resolved
core-jet structure in 5 GHz. B3 1702+457 is clearly resolved with
well defined jet component. The overall radio steep spectrum
indicates that B3 1702+457 is likely a source optically defined as
NLS1 with radio definition of compact steep spectrum sources. From
these three sources, we found that radio loud NLS1s can be either
intrinsically radio loud (e.g. B3 1702+457), or apparently radio
loud due to jet beaming effect (e.g. RXS J16290+4007 and RXS
J16333+4718). 
\end{abstract}





\section{Introduction}

While both permitted and forbidden optical emission lines are
present in narrow line Seyfert 1 galaxies (NLS1), their broad Balmer
lines are narrower than that of normal broad line Seyfert 1 galaxies
with the FWHM of the $\rm H\beta$ line less than 2000 $\rm km
~s^{-1}$ (Osterbrock \& Pogge 1985; Goodrich 1989). Moreover, NLS1
galaxies also exhibit other extreme observational properties, such
as the relatively weak forbidden-line emission, i.e., $\rm [O III]
5007/H\beta < 3$ (Goodrich 1989), strong permitted optical/UV Fe II
emission lines (Boroson \& Green 1992; Grupe et al. 1999;
V\'{e}ron-Cetty et al. 2001), steep soft X-ray spectra (Wang et al.
1996; Boller et al. 1996; Grupe et al. 1998), and rapid X-ray
variability (Leighly 1999; Komossa \& Meerschweinchen 2000). NLS1
galaxies are thought to be young AGNs with relatively small black
hole masses and high accretion rate (indicated by Eddington ratios:
the bolometric-to-Eddington luminosity ratio $L_{\rm bol}/L_{\rm
Edd}\sim1$) (e.g., Boroson 2002; Collin \& Kawaguchi 2004). 

Conventionally, NLS1 galaxies were thought to be radio quiet. With
the discovery of radio loud NLS1s, it is realized that NLS1s simply
have low probability to be radio loud, instead of completely radio
quiet \cite[e.g.][]{kom06}. Compared to about 10\%-15\% for normal
broad line AGNs and quasars (e.g., Ivezi\'{c} et al. 2002), only 7\%
of all NLS1 galaxies are radio loud (Komossa et al. 2006; Zhou et
al. 2006), while only $\sim2.5\%$ of the NLS1s are `very' radio loud
(radio loudness $R>100$). As shown in \cite{kom06}, the radio-loud
NLS1 galaxies are generally compact, steep spectrum sources in the
radio regime, therefore, they are likely associated with compact
steep-spectrum (CSS) radio sources. In contrast, observational
evidences have shown that several of the radio- loudest NLS1
galaxies display blazar characteristics and harbor relativistic jets
\citep{doi07,zho07,yua08}. Yuan et al. (2008) presented a
comprehensive study of a sample of 23 genuine radio-loud NLS1 AGNs
with radio loudness larger than 100. The radio sources of the sample
are ubiquitously compact, unresolved on scales of several
arcseconds. Some of these objects show interesting radio to X-ray
properties that are unusual to most of the previously known
radio-loud NLS1 AGNs, but are reminiscent of blazars, including flat
radio spectra, large-amplitude flux and spectral variability,
compact VLBI cores, very high variability brightness temperatures,
enhanced optical continuum emission, flat X-ray spectra, and
blazar-like spectral energy distributions (SEDs) \cite[also
see][]{fos09a}. Despite that the number of radio loud NLS1s is
rapidly growing \citep{zho06}, mechanisms driving the NLS1 radio
properties is still not clear, such as, accretion rate, black hole
spin, host galaxy properties and merger history.
Although a relativistic jet has been proposed to exist in radio-loud
NLS1s \cite[e.g.][]{kom06,yua08,fos09a}, only until recently its
existence can be strongly proved through the $\gamma-$ray emission
successfully detected in several radio-loud NLS1s
\citep{abd09a,abd09b,abd09c,fos09b}. Especially from these
observations, radio-loud NLS1s were claimed to be a third class of
$\gamma-$ray AGNs, besides blazars and radio galaxies
\cite[e.g.][]{abd09c}. To explore the jet properties in radio-loud
NLS1s, the global e-VLBI observations have been made for the first
$\gamma-$ray detected radio-loud NLS1 PMN J0948+0022 \citep{abd09b},
and more observations are
proposed\footnote{http://www.evlbi.org/gallery/images.html.}.
However in general, still not much is known for the VLBI compact
radio structure of radio-loud NLS1s, from which the jet properties
can be directly explored, such as jet orientation, and jet beaming
effect. Recently, Doi et al. (2006, 2007) performed high-resolution
VLBI observations for several radio-loud NLS1 AGNs and found that
they are unresolved with milliarcsecond resolutions, setting direct
lower limits on the brightness temperatures in the range of
$10^{7}-10^{9}$ K. They also found that inverted radio spectra are
common in the radio-loudest objects. The authors suggested that
Doppler beaming, presumably resulting from highly relativistic jets,
can explain the observations naturally. Most recently, \cite{doi09}
found significant pc-scale structures with high brightness
temperatures for five radio-loud NLS1s from the VLBA high-resolution
observations, indicating nonthermal jets in these sources. Moreover,
the authors claimed that the central engines of some radio loud
NLS1s can generate mildly- or highly-relativistic jets, which make
them apparently radio-loud. However, the nuclei of some NLS1s have
the ability to generate jets significantly strong enough to make
them intrinsically radio-loud.

In this paper, we present the high resolution radio structure of
three radio-loud NLS1s from the unpublished VLBA archive data. The
observations and data reduction is shown in Section 2, and the
results is given in Section 3. The last section is dedicated to
discussions. Throughout the paper, we assume a cosmology with
$H_{0}=71~\rm km~s^{-1}~Mpc^{-1}$, $\Omega_{\rm M}=0.27$, and
$\Omega_{\Lambda}=0.73$ \citep{spe03}. The spectral indices $\alpha$
is defined as $f_{\nu}\propto\nu^{-\alpha}$, in which $f_{\nu}$ is
the flux density at frequency $\nu$.

\section{Observations and Data reduction}


We searched VLBA archive for the unpublished data of the sources
claimed as NLS1s in various literatures. The data of three sources
were finally archived, which covers three radio bands 2.3, 5 and 8.4
GHz. All these three sources have been imaged in 8.4 GHz with
Japanese VLBI Network (JVN) \citep{doi07}, and 1.6 GHz VLBA
observations have been recently shown in \cite{doi09}. The source
list is shown in Table \ref{tab1}, in which the JVN 8.4 GHz flux
density, FIRST 1.4 GHz, GB6 5 GHz flux density and the conventional
radio loudness are given.

All the observations were made in phase referencing mode. The
targets, their corresponding phase referencing calibrator and the
angular distances between target and phase referencing calibrator
are listed in Table \ref{tab2}, most of which lies within a range of
$2.3^\circ$ around the targets except for the observations of B3
1702+457 at two epochs with angular distance of $3.98^\circ$. The
average on-source observational time is about 70 minutes. Data
reductions are made in AIPS. Atmosphere and parallactic angle
effects on data are calibrated before fringe fitting of phase
referencing calibrator are made, and its solutions are applied to
the corresponding target. Bandpass corrections and self-calibrations
are made before data are averaged in 30 seconds so that we can
obtain the results of as high as possible signal to noise ratio. The
imaging and model fitting process is performed in DIFMAP with all
the base contour levels given below $3\sigma$ in the final residual
images. The model fitting results are given in Table \ref{tab3}.




\section{Results and discussion}

\subsection{Brightness temperature}

From the high-resolution VLBA images, the brightness temperature of
radio core $T_{\rm B}$ in the rest frame can be estimated with
\citep{ghi93}

\begin{equation} T_{\rm B}=\frac{S_{\nu}\lambda^2}{2k\Omega_{\rm
s}}=1.77\times10^{12}(1+z)(\frac{S_{\nu}}{\rm Jy})(\frac{\nu}{\rm
GHz})^{-2}(\frac{\theta_{d}}{\rm mas})^{-2} \end{equation} in which
$z$ is source redshift, $S_{\nu}$ is core flux density at frequency
$\nu$, and $\theta_{\rm d}$ is source angular diameter $\theta_{\rm
d}=\sqrt{\rm ab}$ with a and b being major and minor axis,
respectively. The intrinsic brightness temperature $T_{\rm B}^{'}$
can be related with $T_{\rm B}$ by

\be T_{\rm B}^{'}=T_{\rm B}/\delta \label{tb} \ee in which $\delta$
is Doppler factor.

The brightness temperature can also be measured from variability,
which is correspondingly called variability brightness temperature
$T_{\rm B,var}$ and can be estimated as \citep{yua08} \be T_{\rm
B,var}\gtrsim \frac{\Delta P_{\nu e}}{2\pi^{2}k\nu^{2}(\Delta
t)^{2}}=\frac{2D_{\rm L}^{2}\Delta S_{\nu}}{(1+z)\pi k\nu^{2}(\Delta
t)^{2}} \ee where $k$ is the Boltzmann constant, $D_{\rm L}$ the
luminosity distance, $\Delta S_{\nu}$ the variable portion of the
observed flux density, $\Delta P_{\nu e}$ the corresponding radio
power at the emission frequency in the source rest frame, $\nu$ the
observing frequency, and $\Delta t$ the time span in the observer's
frame. The intrinsic brightness temperature $T_{\rm B}^{'}$ can be
related with $T_{\rm B,var}$ by

\be T_{\rm B}^{'}=T_{\rm B,var}/\delta^{3} \label{tbvar} \ee where
$\delta$ is Doppler factor.

Normally, the upper limit of physically realistic brightness
temperature of nonthermal radio emission can be taken as the
equipartition brightness temperature $T_{\rm in}=5\times10^{10}$ K
\citep{rea94}, or the inverse Compton catastrophic brightness
temperature $T_{\rm in}\sim10^{12}$ K (Kellermann \& Pauliny-Toth
1969, but see Singal 2009 for other interpretations). Consequently,
the Doppler factor can be constrained either by equation (\ref{tb})
or equation (\ref{tbvar}). However, it should be noted that the
Doppler factor constrained in these ways is generally smaller than
that from gamma rays \citep{abd09c}. This is due to the fact that
gamma rays are produced in very compact regions, whose compactness
is required for gamma rays to escape, but this in turn means that
these compact blobs are also optically thick to radio frequencies
(synchrotron self-absorption). As the blob is sufficiently expanded
to be optically thin for radio emission, it is likely that it has
decelerated \cite[e.g.][]{bla79}.

\subsection{Individual objects}

\begin{figure}
\epsscale{.75}
\plottwo{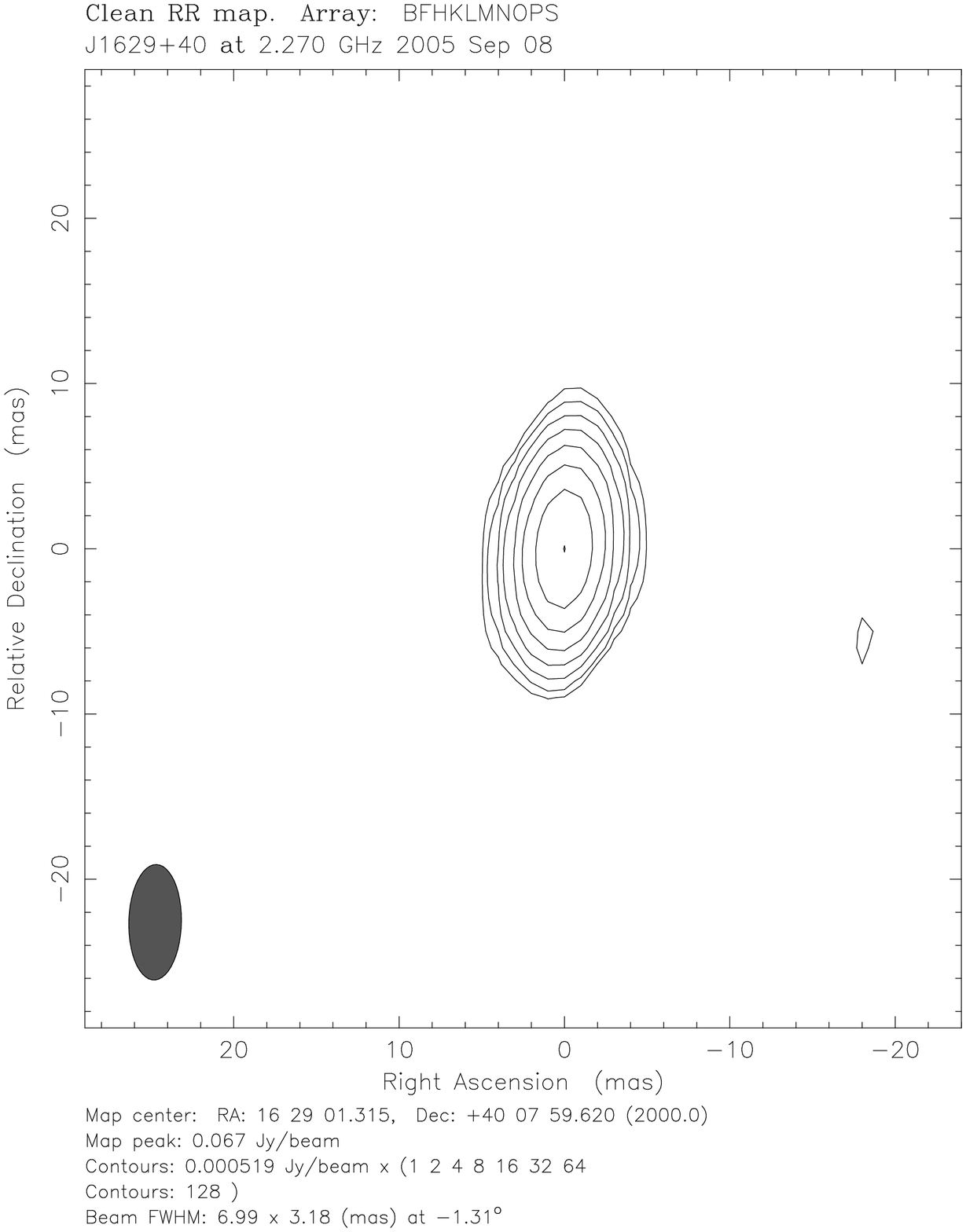}{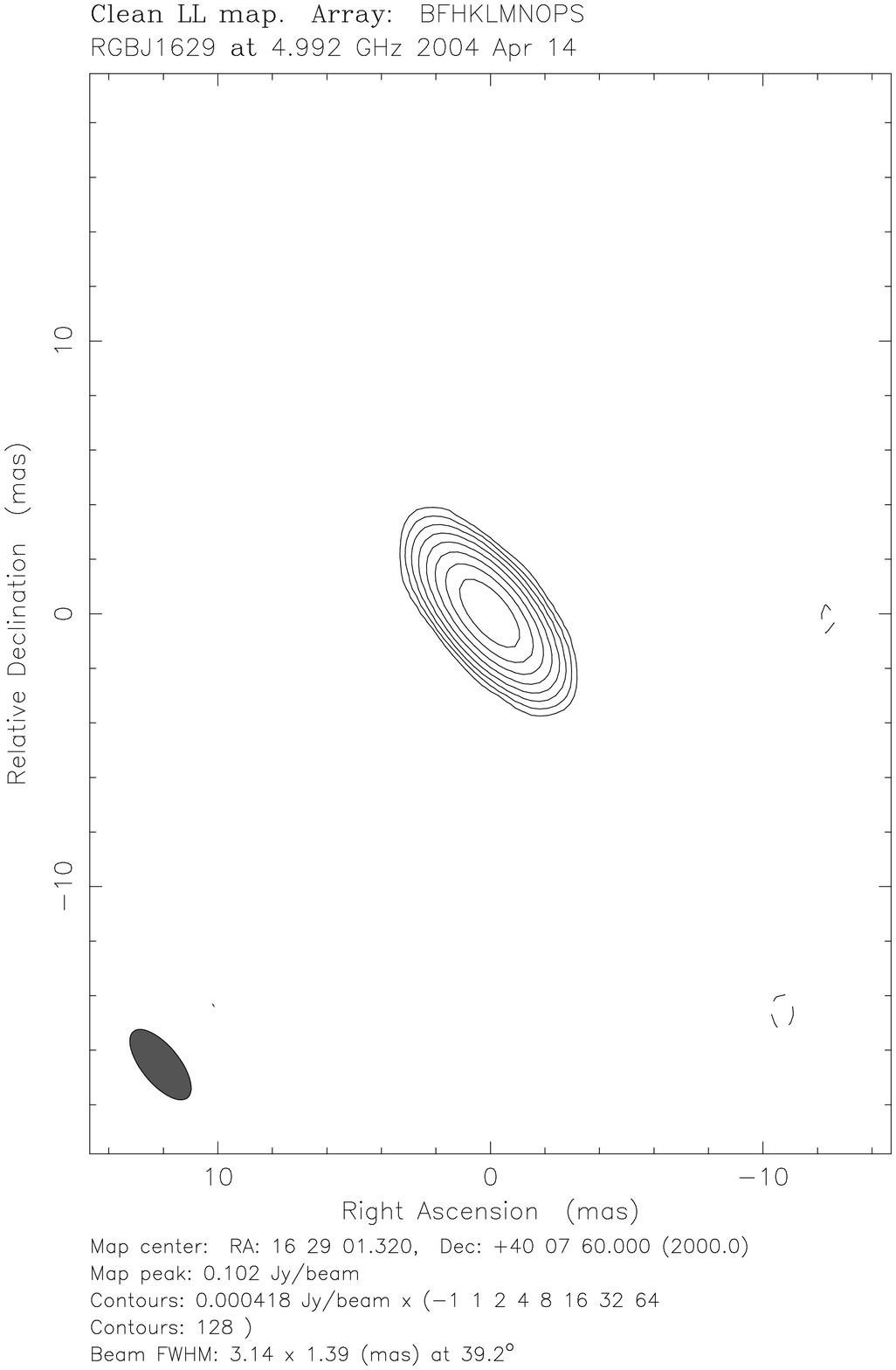}\plottwo{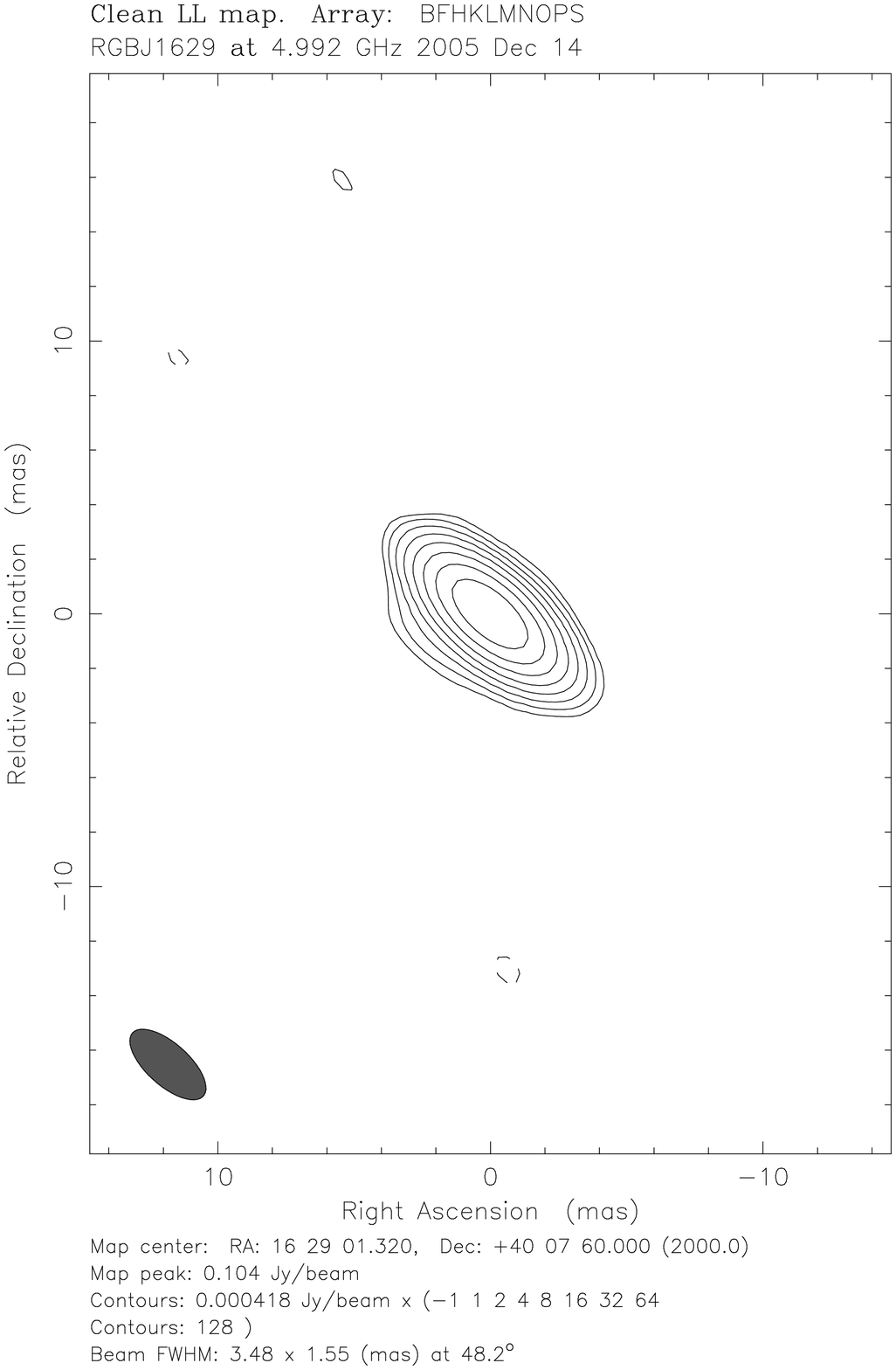}{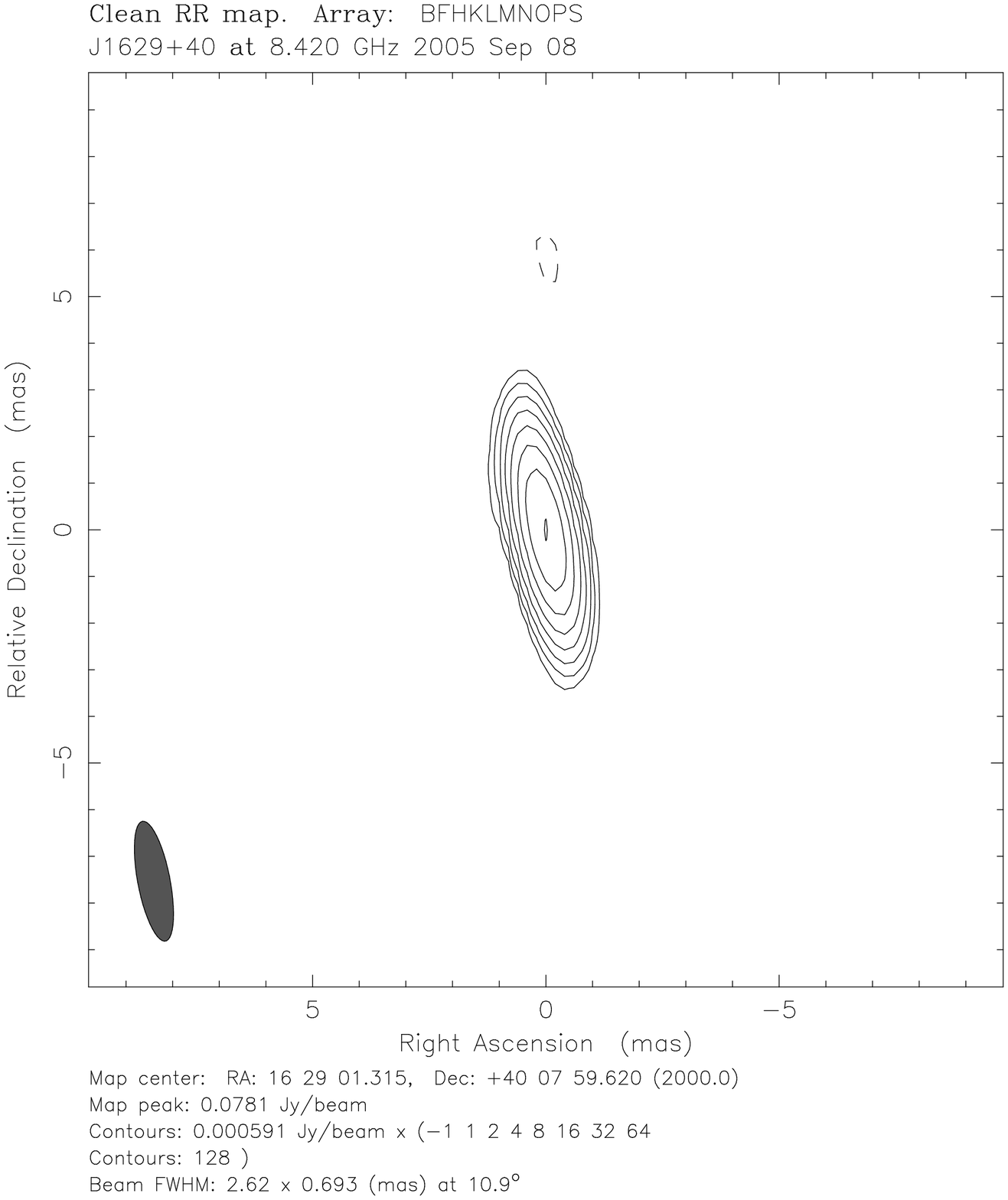}
\caption{The VLBA images of RXS J16290+4007: upper left - 2.3 GHz.
  The restoring beam is 6.99$\times$3.18 mas at P.A. = -1.31$^{\circ}$,
       the contour levels are (1, 2, 4, 8, 16, 32, 64, 128)$\times$0.519 mJy/beam,
       and the peak flux density is 67 mJy/beam; upper right - 5 GHz at Apr. 14, 2004.
  The restoring beam is 3.14$\times$1.39 mas at P.A. = 39.2$^{\circ}$,
       the contour levels are (-1, 1, 2, 4, 8, 16, 32, 64, 128)$\times$0.418 mJy/beam,
       and the peak flux density is 102 mJy/beam; lower left - 5 GHz at Dec. 14, 2005.
  The restoring beam is 3.48$\times$1.55 mas at P.A. = 48.2$^{\circ}$,
       the contour levels are (-1, 1, 2, 4, 8, 16, 32, 64, 128)$\times$0.418 mJy/beam,
       and the peak flux density is 104 mJy/beam;
  lower right: 8.4 GHz. The restoring beam is 2.62$\times$0.69 mas at P.A. = 10.9$^{\circ}$,
       the contour levels are (-1, 1, 2, 4, 8, 16, 32, 64, 128)$\times$0.591 mJy/beam,
       and the peak flux density is 78.1 mJy/beam. \label{fig2}}
\end{figure}

$\it RXS~ J16290+4007$ --- We show the radio structure of this
source in Fig. \ref{fig2}, in which the source is compact and
unresolved in all images, except that it is slightly resolved with
an eastern component at 5 GHz on December 2005. The simultaneous
observations at 2.3 and 8.4 GHz show an inverted spectrum
$\alpha=-0.10$, which is consistent with the inverted spectrum
between FIRST 1.4 and GB6 5 GHz $\alpha=-0.28$ \cite[see Table
\ref{tab1}, see also][]{zho02}. The brightness temperature at all
wavebands are high $T_{\rm b}\gtrsim 10^{11}~\rm K$, especially it
reaches $T_{\rm b}= 10^{12.4}~\rm K$ at 8.4 GHz. The inverted
spectrum, compact structure and high brightness temperature all
together suggest a Doppler beaming effect in this source. This is
further supported by the flux variations. Although the variation is
not evident during two epoch 5 GHz observations (see Table
\ref{tab3}), large variations are found when comparing with the
previous data at 5 GHz \cite[22 mJy,][]{ver01} and 8.4 GHz \cite[JVN
26.3 mJy,][]{doi07}. At 8.4 GHz, the variability brightness
temperature is quite high $T_{\rm b,var}=10^{12.2}$ K (see Table
\ref{tab3}). The high $T_{\rm B}$ values are commonly explained as
emission originating from relativistic jets \citep{bla79}.
Conservatively taking the inverse Compton limit $10^{12}~ \rm K$,
the minimum Doppler factor can be estimated as $\delta_{\rm
min}=T_{\rm b}/10^{12}$ or $\delta_{\rm min}=(T_{\rm
b,var}/10^{12})^{1/3}$. This results in $\delta_{\rm min}=1.2-2.3$,
which can be larger if the equipartition brightness temperature
$5\times10^{10}$ K is used as limit \citep{rea94}. To conclude, this
source can be a blazar-like NLS1s with jet pointing towards us with
small viewing angle \cite[see also][]{doi07,yua08,doi09}, which is
consistent with the results presented in \cite{pad02}.

\begin{figure}
\epsscale{.7} \plotone{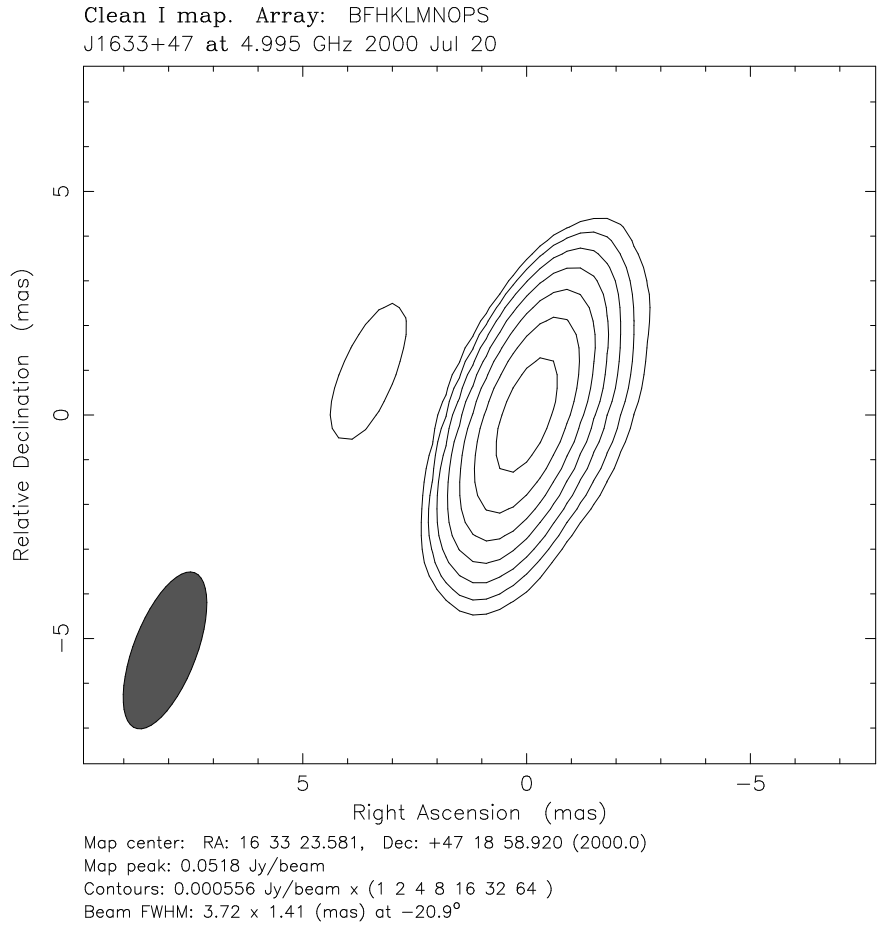} \caption{The VLBA 5 GHz image of
RXS J16333+4718.
  The restoring beam is 3.72$\times$1.41 mas at P.A. = -20.9$^{\circ}$,
       the contour levels are (1, 2, 4, 8, 16, 32, 64)$\times$0.556 mJy/beam,
       and the peak flux density is 51.8 mJy/beam.\label{fig3}}
\end{figure}

\begin{figure}
\epsscale{1.} \plottwo{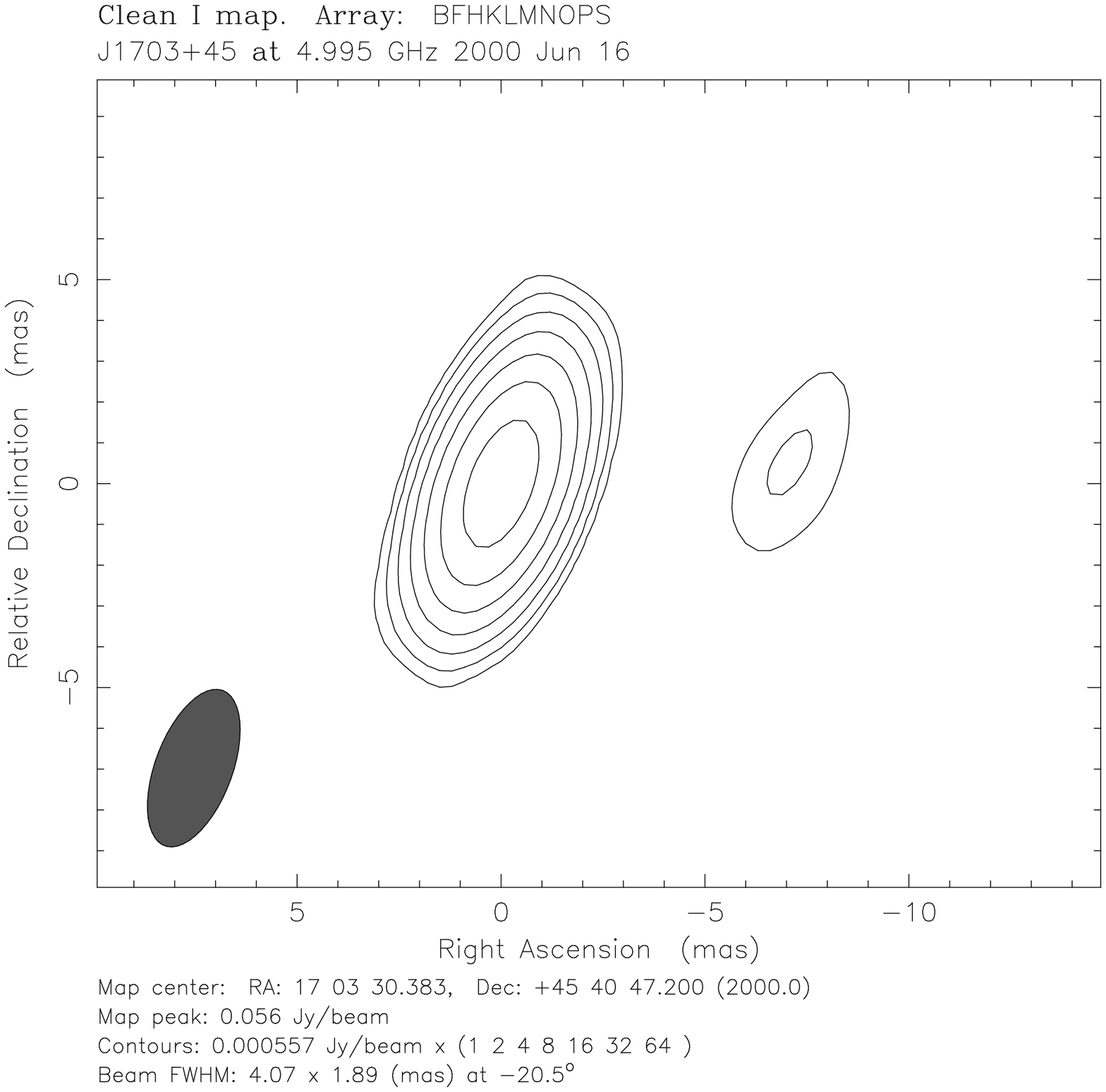}{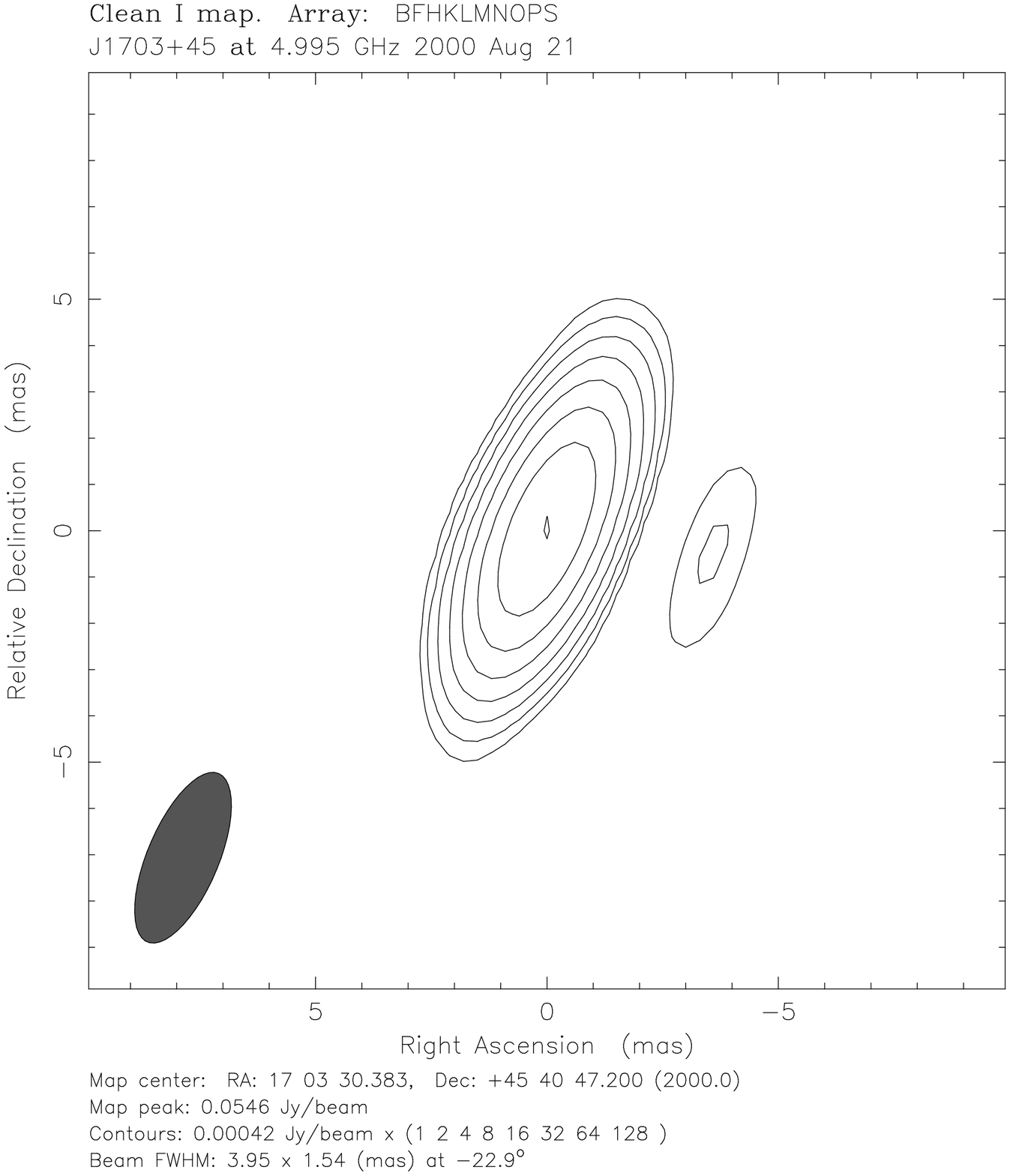} \caption{The VLBA 5 GHz
images of B3 1702+457: left - at Jun. 16,
  2000. The restoring beam is 4.07$\times$1.89 mas at P.A. = -20.5$^{\circ}$,
       the contour levels are (1, 2, 4, 8, 16, 32, 64)$\times$0.557 mJy/beam,
       and the peak flux density is 56 mJy/beam;
       right - at Aug. 21, 2000. The restoring beam is 3.95$\times$1.54 mas at P.A. = -22.9$^{\circ}$,
       the contour levels are (1, 2, 4, 8, 16, 32, 64, 128)$\times$0.42 mJy/beam,
       and the peak flux density is 54.6 mJy/beam. \label{fig4}}
\end{figure}

Tentatively putting our data at 2.3, 5 and 8.4 GHz together, we find
the radio spectrum are resemble to that of High Frequency Peakers
(HFPs), i.e. compact objects with a simple convex radio spectrum
turning over at frequencies well above a few GHz, which likely
represent the earliest stage in individual radio source evolution
\cite[e.g.][]{dal00,ori09}. While this should be checked with
further simultaneous multiwaveband observations, as a matter of
fact, its spectral shape above 5 GHz is inverted, as measured by
simultaneous multiwavelength observations with Effelsberg 100-m
telescope by \cite{neu94}. Alternatively, it is likely that the
beaming effect boosted the radio emission as well as the peak
frequency of the synchrotron self-absorption spectra
\cite[e.g.][]{doi07}. As pointed out by \cite{ori09}, HFPs are
selected on the basis of both the convex shape of their spectra and
the high frequency at which the spectral peak occurs. In this way,
it is possible that boosted objects matching the selection criteria
contaminate the HFPs sample. Indeed, from the analysis of the radio
properties, \cite{ori09} found that HFPs galaxies are not variable,
are unpolarized and exhibit a Double/Triple structure, typical of
young radio sources. On the other hand, the majority of HFPs quasars
are strongly variable, polarized and with Core-Jet structure, as
expected in beamed objects. Therefore, RXS J16290+4007 is likely a
boosted object rather than a genuine HFPs young radio sources.
\\
\\
$\it RXS~ J16333+4718$ --- This source is unresolved at JVN 8.4 GHz
milliarcsecond resolution (Doi et al. 2007), and at 1.6 VLBA GHz
\citep{doi09}. Its spectral shape above 5 GHz is inverted, as
measured by simultaneous multiwavelength observations by Neumann et
al. (1994). We present 5 GHz radio structure in Fig. \ref{fig3}, at
which it is slightly resolved into core-jet structure. Combining JVN
8.4 GHz \citep{doi07} and VLBA 1.6 GHz data \citep{doi09} with our 5
GHz data, we find that the core spectral index is flat between 1.6
and 5 GHz $\alpha=0.30$, which is consistent with that of
\cite{zho02} (see Table \ref{tab1}). The high brightness temperature
at 5 GHz $T_{\rm b}=10^{11.3}$ K in combination with the flat
spectrum indicate that an at least mild relativistic jets is likely
responsible for the radio emission and structure. However, the
spectra is steep between 1.6 and 8.4 GHz $\alpha=0.73$
\citep{doi09}, and between 5 and 8.4 GHz $\alpha=1.73$. The steep
spectrum and compact radio structure make this source resemble to
CSS source. This however is not conclusive, since the observations
were not performed simultaneously. Indeed, the variation can be
clearly seen when comparing GB6 with VLBA 5 GHz data. Therefore,
further simultaneous multi-epoch VLBI observations are needed to
investigate the spectral index of this source, then to explore the
source nature.
\\
\\
$\it B3~ 1702+457$ --- This source was classified as Compact Steep
Spectrum (CSS) sources with turnover frequency $\nu_{\rm peak}<150$
MHz in the sample of Compact Radio sources at Low Redshift (CORALZ)
\citep{sne04}. The VLBA 5 GHz radio structure is shown in Fig.
\ref{fig4}, in which the source is resolved into core-jet structure
with weak radio jet components clearly seen in two images. The
spectral index between FIRST 1.4 GHz and GB6 5 GHz show a steep
spectrum with $\alpha=1.23$ (see Table \ref{tab1}).
Consistently, the VLBA 1.6 GHz \citep{doi09} and JVN 8.4 GHz
\citep{doi07} show a steep index $\alpha=0.84$, which is also
present between VLBA 5 GHz and JVN 8.4 GHz $\alpha=2.12$ (see Tables
\ref{tab1} and \ref{tab3}). In contrast, the spectral index of the
prominent radio component become flat $\alpha\sim0.26$ with VLBA 1.6
and 5 GHz data, from which the radio core can be identified.
However, VLBA 1.6 GHz and 5 GHz were not obtained simultaneously,
therefore, the steep spectrum of 1.6 GHz and 5 GHz can not be
completely excluded due to the variability. Indeed, the flux
variation is apparent at 5 GHz. The GB6 flux density is 24.7 mJy
(see Table \ref{tab1}), whereas the flux density of VLBA core is
56.8 mJy on June 2000 (see Table \ref{tab3}). It thus is possible
that the VLBA 5 GHz data were taken when the source was at high
state, while 1.6 GHz data was at relatively low state, which results
in a flat spectrum between VLBA 1.6 GHz and 5 GHz. The brightness
temperature of this source does not show a severe beaming effect,
which is not inconsistent with the steep spectrum nature of this
source. In order to investigate the genuine spectral index, it is no
doubt that the simultaneous observations are required to explore the
nature of the radio structure. In addition, the flux monitoring also
enable us to explore the nature of this source since genuine CSS
sources usually do not show voilent variations \citep{ode98}.

\section{Discussion}


While the high brightness temperature and compact radio structure
indicate the jet origin of radio emission, the spectral index are
not homogenous in three sources. The inverted spectrum of RXS
J16290+4007 from simultaneous 2.3 and 8.4 GHz observations has shown
that beaming effect plays important role in making high radio
loudness, which is supported by the inverted spectrum above 5 GHz
measured by simultaneous multiwavelength observations \citep{neu94}.
Similarly, the beaming effect may likely also present in RXS
J16333+4718, as shown by the inverted spectrum above 5 GHz
\citep{neu94}. This is supported by the flat spectrum between 1.6
and 5 GHz, despite that further simultaneous multi-band observations
are needed for confirmation. These two sources are thus likely
intrinsically radio quiet or intermediate (depends on the strength
of beaming effect), however, are apparently radio loud due to the
beaming effect in radio emission \cite[see also][]{doi09}. In
contrast, the overall spectrum of B3 1702+457 are steep, in
combination with the compact structure, making this source resemble
to CSS sources. This source can be intrinsically radio loud since
the beaming effect may not be important. Indeed, the radio loudness
of this source is only $\rm R=11$, relatively small compared to RXS
J16290+4007 ($\rm R=182$) and RXS J16333+4718 ($\rm R=205$, see
Table \ref{tab1}). To summarize, radio loud NLS1s can be either
intrinsically radio loud, or apparently radio loud due to jet
beaming effect \cite[see also][]{doi09}.

Both having strong emission lines and Doppler boosted jet emission,
the most radio loud NLS1s and flat spectrum radio quasars are
presumably related together. Indeed, several radio-loudest NLS1
galaxies have been found to display blazar characteristics and
harbor relativistic jets \citep{doi07,zho07,yua08}, which have been
strongly proved by the detected $\gamma-$rays
\citep{abd09a,abd09b,abd09c,fos09b}. As claimed by \cite{yua08}, in
a sample of 23 radio-loud NLS1 AGNs with radio loudness larger than
100, some objects show radio to X-ray properties that are
reminiscent of blazars, some of which even resemble
high-energy-peaked flat spectrum radio quasars (HFSRQs) in their
SEDs with synchrotron peak frequency likely at around UV/X-ray
regimes. RXS J16290+4007, the Doppler boosted object as shown in
Section 3.2, was suggested by \cite{pad02} to be the first HFSRQ for
its modeled synchrotron peak at $2\times10^{16}$ Hz. A steep soft
X-ray spectrum with photon index $\Gamma\simeq2.5$ implies a
synchrotron origin for X-ray emission \citep{pad02,gra06}. While
HFSRQs represent the outlier of blazar sequence \citep{fos98,ghi98},
they have not definitely been found yet \citep{lan08,mar08}.
Nevertheless, the existence of HFSRQs is explained in \cite{ghi08},
in which the jet dissipation region is out of the broad-line region,
resulting in a much reduced energy density of BLR photons in the jet
region. Therefore, the cooling due to the inverse Compton process is
not severe, causing a high synchrotron peak frequency although the
source luminosity can be high. Taking the disk-corona emission into
account and assuming the jet dissipation out of BLR, the jet
parameters of RXS J16290+4007 has been estimated by \cite{mar08}
through SED modeling, of which the X-ray emission are dominated by
the synchrotron emission from jet. Although it is regarded as
HFSRQs, the authors found that RXS J16290+4007 well follows the
general blazar spectral sequence established by blazars in
\cite{cel08} in terms of parameter diagram of $\gamma_{\rm peak}$,
the energy of electrons radiating the peak synchrotron luminosity,
and the radiation energy density (magnetic field plus synchrotron
photons) or the total jet power \cite[see their Figs. 9 and 11
in][]{mar08}.

In another blazar-like NLS1, RXS J1633+4718, a flat X-ray spectrum
with $\Gamma=1.37\pm0.49$ found by \cite{yua08}, is not inconsistent
with the typical X-ray spectra of FSRQs, which are flat and extend
from the soft to the hard X-ray band ($\Gamma=1.6$ with a dispersion
$\sigma= 0.1-0.2$; see, e.g., Reeves \& Turner 2000). Such flat
X-ray spectra is usually interpreted as inverse Compton radiation
from the relativistic jets and beaming. On the other hand, the
prominent soft X-ray excess is also found in RXS J1633+4718, which
is typical of normal NLS1 galaxies \citep{yua08}. Intriguingly,
however, RXS J1633+4718 shows a broadband SED similar to that of
HFSRQs \citep{yua08}. For B3 1702+457, the CSS-like source, the
spectral index at 2-10 keV is $\Gamma=2.20\pm0.06$, and the soft
X-ray excess is evident, which is typical of normal NLS1 galaxies
\citep{vau99}. It might be true that the X-ray emission of radio
loud NLS1s can be a mixture of the emission from disk-corona and
nonthermal jets. The fraction of each components may alter with the
significance of beaming effect, in the way that the disk-corona
emission may dominate over nonthermal one for sources with little
(if any) beaming effect (e.g. B3 1702+457; PKS 2004-447 in Gallo et
al. (2006) and Foschini et al. (2009a), but see Abdo et al. (2009c)
for alternatives), however, nonthermal one may dominate in case of
significant beaming effect \cite[e.g. RXS J16290+4007,
see][]{mar08}. The similar competition was found in FSRQs from the
optical variability in the form that the nonthermal jet emission
becomes dominant when source is brighter, however, thermal emission
from accretion disk is dominant when source is at low state,
resulting in a tendency of redder when brighter \cite[e.g.][]{gu06}.
Although the beaming effect are usually believed to be nontrivial in
FSRQs, the thermal emission from accretion disk can be significant
and even are dominant in the optical region
\citep[e.g.][]{rai07,che09}. Interestingly, the multiwavelength
campaign of the $\gamma-$ray detected radio-loud NLS1 PMN J0948+0022
from radio to $\gamma-$rays, shown the emerging of the accretion
disk emission when the synchrotron emission decreased
\citep{abd09b}. It is likely also the case in X-ray waveband
\cite[e.g.][]{fos09a}, though it is claimed that in most radio loud
quasars the contribution from a hot disk corona to the observed
X-rays is negligible (in the hard X-ray band $>$2 keV), except for
the steep-spectrum soft X-ray excess below 1 keV (e.g., Brinkmann et
al. 1997; Yuan et al. 2000). Indeed, the integrated model including
synchrotron jet emission, disk-corona emission and inverse Compton
emission from jet has been used to model SED of NLS1s, from which
the nature of radio loud NLS1s can be well studied
\citep{abd09a,abd09b,abd09c,fos09b}.

While it strongly confirms the presence of a relativistic jet in
radio-loud NLS1s, the $\gamma-$ray detection is also important to
study the jet properties by modeling the SEDs, e.g. the jet power,
from which the characteristic of radio-loud NLS1s can be explored.
Through the model fit including synchrotron self-Compton (SSC),
external Compton (EC), and accretion disk-corona in four gamma-ray
detected NLS1s, their jet powers are found in the average range of
blazars with two sources in the region of quasars, and another two
in the range of typical of BL Lac objects \citep{abd09c}. However,
the main differences with respect to blazars are in the black hole
masses and accretion rates, as argued in \cite{abd09c}, with the
former about 1-2 orders of magnitude lower than the typical blazar
masses, and the later obviously higher than those of blazars.
Moreover, blazars are usually hosted by elliptical galaxies, while
it is likely to be spiral ones in radio-loud NLS1s
\cite[e.g.][]{zho06}. From these observational evidences,
\cite{abd09c} claimed that radio-loud NLS1s may represent a third
subset of $\gamma-$ray AGNs, besides blazars and radio galaxies. If
this is the case, it remains unclear whether radio-loud NLS1s should
follow the blazar sequence (e.g. RXS J16290+4007 in this work). This
certainly needs further investigations. On the other hand, it is
still not clear why radio-loud NLS1s host a relativistic jet, and
how it is formed. It can be even more complicated in terms of the
fact that the host galaxies of NLS1s is generally of spiral type,
which breaks the paradigm associating relativistic jets with giant
elliptical \cite[e.g.][]{mars09}.

Although the accretion disk and jet are found to be closely related
\cite[e.g.][]{cao01,gu09a}, the details of disk-jet coupling is not
known yet, besides that jet formation and radio loud/quiet dichotomy
of AGNs are not well understood \cite[e.g.][]{tch09}. As one
possibility, jet activity can be intermittent, due to, for example,
the accretion disk instability \cite[e.g.][]{cze09}, which is
recently adopted to explain CSS/GPS sources \citep{wu09a}. The
existence of double double radio sources seems to support the
intermittent scenario \cite[e.g.][]{mar09}. Optically, NLS1s are
thought to be young AGNs with small black hole mass accreting at
high accretion rate, implying the central accretion process are at
the early stage of accretion history. In radio band, the compact
nature of radio structure of CSS sources are believed to be due to
the fact that sources are at early stage of evolutionary history, at
which radio jets do not have time to expand to large scale yet
\cite[e.g.][]{ode98,gu09b,wu09b}. The high occurrence of sources
optically defined as NLS1s with radio definition of CSS or GPS,
suggests that the connection could have a physical origin
\citep{gal06,kom06}, for example, perhaps the black hole environment
and the radio component are forming simultaneously. It has been
proposed that the narrow-line emission nebula of GPS/CSO galaxies is
likely still in the early phases of its evolution, meanwhile they
are at early evolutionary phases of radio-loud AGN \citep{vin06}. It
might also be interesting to note that low-power CSS can probably be
the parent population of the flat-spectrum RL NLS1 galaxies with
beamed jet emission, as speculated in \cite{yua08}. Therefore,
NLS1-CSS/GPS sources, for example, B3 1702+457 in this work, can be
precious targets to explore the triggering of accretion process as
well as the jet activity. As an example, the investigation of the
SED of PKS 2004-447, a CSS and possible radio-loud NLS1, was
recently presented by \cite{abd09c}, utilizing radio to $\gamma-$ray
data. Although the scarce, non-simultaneous data and the weakness of
the $\gamma$-ray emission, preclude the tightly constrains on jet
parameters, the jet power of PKS 2004-447 is claimed to be in the
range typical of BL Lac Objects.
As it may give
clues on the trigger of AGNs activity, the potential connection
between radio-loud NLS1 and CSS/GPS sources requires further
investigations based on the large sample of radio-loud NLS1s
\cite[e.g.][]{zho06}.

\acknowledgments

We thank the anonymous referee for insightful comments and
constructive suggestions, which were greatly helpful in improving
our paper. We thank Weimin Yuan, Akihiro Doi and Biping Gong for
helpful discussions. This work is supported by National Science
Foundation of China (grants 10633010, 10703009, 10821302, 10833002
and 10978009), by Science and Technology Commission of Shanghai
Municipality (09ZR1437400), by Scientific Research Foundation for
Returned Scholars, Ministry of Education of China (9020090306) and
by 973 Program (No. 2009CB824800). This research has made use of the
NASA/ IPAC Extragalactic Database (NED), which is operated by the
Jet Propulsion Laboratory, California Institute of Technology, under
contract with the National Aeronautics and Space Administration.



{\it Facilities:} \facility{VLBA}

\clearpage



\clearpage

\clearpage


\clearpage

\begin{table*}\centering\begin{minipage}{160mm}
\caption{The source list.\label{tab1}}
\begin{tabular}{lcccccc}
\hline\hline
 Source & z & $S_{\rm 1.4 GHz}^{\rm FIRST}$ (mJy) & $S_{\rm 5 GHz}^{\rm GB6}$ (mJy) & $S_{\rm 8.4 GHz}^{\rm JVN}$ (mJy) & $R_{\rm 5 GHz}$ & $\alpha_{r}$  \\
\hline
RXS J16290+4007 &  0.272  &  11.9         & 17.0  & 26.3  & 182 & -0.28 \\
RXS J16333+4718 &  0.116  &  65.0         & 33.6  & 21.2  & 205 & 0.52  \\
B3 1702+457     &  0.061  &  118.6        & 24.7  & 18.5  & 11  & 1.23  \\
\hline
\end{tabular}
\begin{quote}
Column (1): Source name; Column (2): redshift; Column (3): FIRST 1.4
GHz flux density (Becker et al. 1995); Column (4): GB6 5 GHz flux
density \citep{gre96}; Column (5): JVN 8.4 GHz flux density
\citep{doi07}; Column (6): radio loudness in Zhou \& Wang (2002);
Column (7): spectral index between 1.4 and 5 GHz.

\end{quote}
\end{minipage}
\end{table*}

\clearpage

\begin{table*}\centering\begin{minipage}{160mm}
\caption{VLBA archive data.\label{tab2}}
\begin{tabular}{lcccccc}
\hline\hline
 Source & z & $\nu$ (GHz) & Date & ID & Calibrator & Dis. (deg) \\
\hline
RXS J16290+4007 &  0.272  &  5         &  04/14/2004  &  BP109A  & J1625+4134  & 1.56  \\
                &         &  5         &  12/14/2005  &  BP123C  & J1625+4134  & 1.56  \\
                &         &  2.3, 8.4  &  09/08/2005  &  BE042E  & J1640+3946  & 2.23  \\
RXS J16333+4718 &  0.116  &  5         &  07/20/2000  &  BM133C  & J1631+4927  & 2.17  \\
B3 1702+457     &  0.061  &  5         &  06/16/2000  &  BM133D  & J1713+4916  & 3.98  \\
                &         &  5         &  08/21/2000  &  BM133E  & J1713+4916  & 3.98  \\
\hline
\end{tabular}
\begin{quote}
Column (1): Source name; Column (2): redshift; Column (3):
frequency; Column (4): observational date; Column (5): Program ID;
Column (6): phase referencing calibrator; Column (7): angular
distance of phase referencing calibrator to object in unit of
degree.
\end{quote}
\end{minipage}
\end{table*}

\clearpage

\begin{deluxetable}{lcccccccccc}
\tabletypesize{\scriptsize} \rotate \tablecaption{Observational
results.\label{tab3}} \tablewidth{0pt} \tablehead{ \colhead{Source}
& \colhead{$\nu$} & \colhead{Date} & \colhead{Comps.} &
\colhead{$r$} & \colhead{$\theta$} & \colhead{$S$} & \colhead{$a$} &
\colhead{$b/a$} & \colhead{log $T_{\rm B}$} & \colhead{log $T_{\rm
B,var}$} \\
   \colhead{} & \colhead{(GHz)} & \colhead{}  & \colhead{} & \colhead{(mas)} & \colhead{(deg)}
    & \colhead{(mJy)} & \colhead{(mas)} & \colhead{} & \colhead{(K)} & \colhead{(K)} } \startdata
RXS J16290+4007 &   5         &  04/14/2004 & C  & ...   & ...     & 104.9 & 0.45  & 0.61 &  10.9 & 9.9     \\
                &   5         &  12/14/2005 & C  & ...   & ...     & 107.8 & 0.50  & 0.54 &  10.9 & ...     \\
                &             &             & E  & 1.55  & 121.63  & 2.2   & 1.25  & 1.00 &  ...  & ...     \\
                &   2.3       &  09/08/2005 & C  & ...   & ...     & 68.9  & 0.56  & 1.00 &  11.0 & ...     \\
                &   8.4       &  09/08/2005 & C  & ...   & ...     & 78.1  & 0.03  & 1.00 &  12.4 & 12.2    \\
RXS J16333+4718 &   5         &  07/20/2000 & C  & ...   & ...     & 51.9  & 0.15  & 1.00 &  11.3 & 8.7     \\
                &             &             & E  & 3.67  & 79.15   & 0.9   & 0.28  & 1.00 &  ...  & ...     \\
                &             &             & W  & 1.24  & -114.43 & 2.4   & 3.74  & 0.09 &  ...  & ...     \\
B3 1702+457     &   5         &  06/16/2000 & C  & ...   & ...     & 56.8  & 0.56  & 0.32 &  10.6 & 8.4     \\
                &             &             & W  & 7.28  & -85.20  & 1.5   & 1.13  & 1.00 &  ...  & ...     \\
                &   5         &  08/21/2000 & C  & ...   & ...     & 55.6  & 0.33  & 0.37 &  11.0 & ...     \\
                &             &             & W  & 3.60  & -94.97  & 1.0   & 0.07  & 1.00 &  ...  & ...     \\

\enddata
\tablecomments{Column (1): Source name; Column (2): observing
frequency; Column (3): observing date; Column (4): components: C =
core, E = eastern components and W = western components; Column (5):
the distance to core; Column (6): position angle; Column (7): flux
density; Column (8): major axis; Column (9): axial ratio; Column
(10): brightness temperature; Column (11): variability brightness
temperature.}
\end{deluxetable}




\end{document}